\documentclass[a4paper]{article}

\usepackage{INTERSPEECH2019}
\usepackage[colorlinks,linkcolor=blue]{hyperref}
\usepackage{stfloats}

\title{ShaneRun System Description to VoxCeleb Speaker Recognition Challenge 2020}
\name{Shen Chen}
\address{Delta Electronics}
\email{chenshen@next-aiot.cn}

\begin{document}

% title
\maketitle

% abstract
\begin{abstract}
In this report, we describe the submission of ShaneRun's team to the VoxCeleb Speaker Recognition 
Challenge (VoxSRC) 2020. We use ResNet-34 as encoder to extract the speaker embeddings, 
which is referenced from the open-source voxceleb-trainer. 
We also provide a simple method to implement optimum fusion using t-SNE normalized distance of 
testing utterance pairs instead of original negative Euclidean distance from the encoder. 
The final submitted system got 0.3098 minDCF and 5.076 \% ERR for Fixed data track, 
which outperformed the baseline by 1.3 \% minDCF and 2.2 \% ERR respectively.
\end{abstract}

% index term
\noindent\textbf{Index Terms}: speaker recognition, VoxCeleb, t-SNE, optimum fusion

% introduction
\section{Introduction}
The VoxCeleb Speaker Recognition Challenge 2020 is second term of the new series of 
speaker recognition challenges. The goal of this challenge is to probe how well current methods can 
recognize speakers from speech obtained 'in the wild' \cite{Chung19-V2T}.   
This year, the challenge has four separate tracks: Speaker verification - Fixed training data, 
Speaker verification - Open training data, Speaker verification - Self-supervised and 
Speaker diarisation - Open training data. Our team only participated the track of 
Speaker verification - Fixed training data. In the Fixed condition, 
participants can only use the development part of VoxCeleb2 as training data, 
including 1,092,009 utterances from 5,994 speakers \cite{Chung18-VDS}.
The following sections of this report describe the details of our submission to this challenge.

% proposed method
\section{System design}
\subsection{ResNet-34 based encoder}
We use 64-dimensional log Mel-filter banks - 16kHz, with a hamming window of 25 ms width and 10 ms step.
The segment length used for training is fixed 2 seconds which is extracted randomly from each utterances.
We build our encoder based on the open-source voxceleb-trainer which is built based on ResNet-34 \cite{He16-DRL}, 
and is available to download at github\footnote{\url{https://github.com/clovaai/voxceleb_trainer}}.
The overview table of our encoder is shown in Table~\ref{tab:encoder}. 
The detailed topology of used ResNet-34 is the same as \cite{Chung20-IDO}.

\begin{table}[t]
  \caption{Overview of encoder.}
  \label{tab:encoder}
  \centering
  \begin{tabular}{lll}
    \toprule
    \textbf{Category} & \textbf{Item} & \textbf{Value}              \\
    \midrule
    acoustic          & feature       & 64-d Mel-filter bank        \\
    acoustic          & sampling rate & 16 kHz                      \\
    acoustic          & segment length& 2 seconds                   \\
    data              & pre-processing& No                          \\
    data              & augmentation  & additive noise and RIR\cite{Heo20-CBS}  \\
    frontend          & topology      & ResNet-34 + relu + BN       \\
    frontend          & loss function & softmax + prototypical\cite{Chung20-IDO}   \\
    frontend          & pooling       & SAP\cite{Chung20-IDO}                      \\
    frontend          & embedding dim.& 512                         \\
    backend           & scoring       & negative Euclidean distance \\
    \bottomrule
  \end{tabular}
\end{table}

We build several systems to get scores output for validation set and voxsrc2020 test set as
listed in Table~\ref{tab:systems}.
The only difference between Fast ResNet-34 and Half ResNet-34 is the channel depth, 
wherein Fast ResNet-34 is one fourth of original ResNet-34 and Half ResNet-34 is half of it.

For system-1, we train it in single GPU (GeForce GTX 1080 TI), and it takes about 12 minutes for 
one epoch, and about 5 days to run 500 epochs. For system-2 and system-3, we use the pre-trained model 
from baseline. The eval. Frame of 0 means using full utterance length without random extraction, 
and augmentation means data augmentation.

\begin{table}[t]
  \caption{Overview of systems.}
  \label{tab:systems}
  \centering
  \begin{tabular}{llll}
    \toprule
    \textbf{System} & \textbf{Topology} & \textbf{Eval. frame}  & \textbf{Augmentation}  \\
    \midrule
    system-1        & Fast ResNet-34    & 4 seconds             & No                     \\
    system-2        & Half ResNet-34    & 0 or 4 seconds        & No                     \\
    system-3        & Half ResNet-34    & 0 or 4 seconds        & Yes                    \\
    \bottomrule
  \end{tabular}
\end{table}

\subsection{t-SNE normalization}
't-Distributed Stochastic Neighbor Embedding' or 't-SNE' is a kind of technique aims to 
alleviate "crowding problem" \cite{Maaten08-VDU}. Similar to "SNE" \cite{Hinton02-SNE}, t-SNE can convert the high-dimensional 
Euclidean distances between data points into conditional probabilities that represent similarities.

\subsection{Optimum fusion}
The 'BOSARIS Toolkit' in \cite{Brummer13-TBT} is one kind of commonly used toolkit for system fusion. 
In this paper, we try optimum fusion by Python script, wherein the input are original scores files and 
the output is optimum weight and fused scores.

% experiments
\section{Experiments}
\subsection{Datasets} 
We use development part of VoxCeleb2 dataset for training, which contains 5,994 speakers. 
The VoxCeleb1 test sets \cite{Nagrani17-VAL} are used as validation sets.

\subsection{Training}
\begin{figure}[t]
  \centering
  \includegraphics[width=\linewidth]{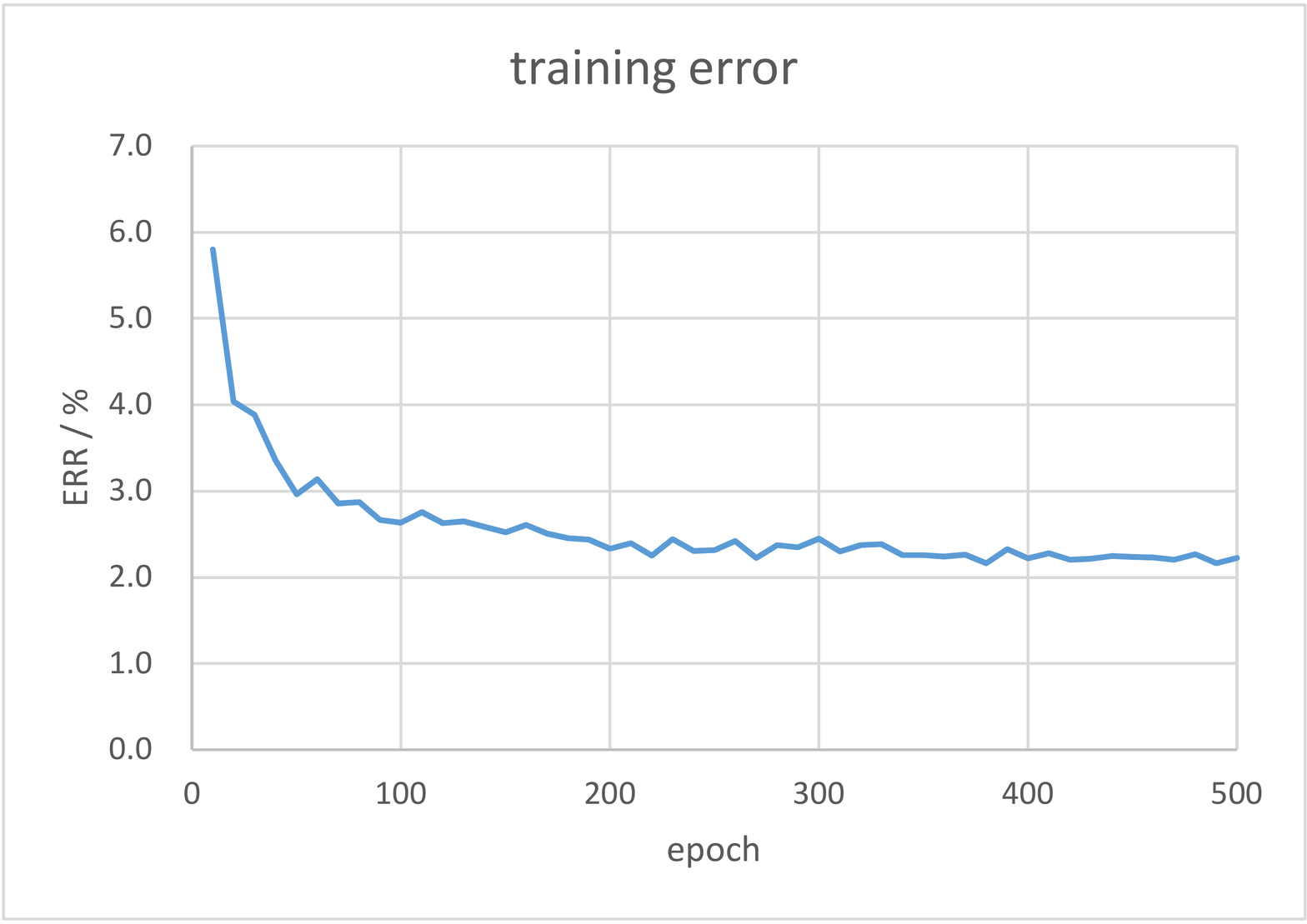}
  \caption{Training curve.}
  \label{fig:itercurve}
\end{figure}

The network is trained in PyTorch using a single GPU (GeForce GTX 1080 TI) with adam optimizer, 
and run for 500 epochs.   
Following \cite{Chung20-IDO}, we use a initial learning rate with 0.001, and rate decay of 0.95 every 10 epochs. 
The batch size is set to 270 to make full use of GPU memory (11 GB).   
The iteration curve of training system-1 is shown in Figure~\ref{fig:itercurve}. 
The validation error is printed every 10 epochs during training. 
Therefore, the starting error is approximately 6.0 \% at the tenth epoch.
% figure

\subsection{Results}
The validation results are shown in Table~\ref{tab:example}. The single best scores is system-2 with 4-second 
evaluation frame which is highlighted in bold.   
The 'fusion-avg' term is average fusion of scores-2, scores-3, scores-4 and scores-5, 
while 'fusion-opt' term is optimum fusion of them.   
The final submitted system used optimum fusion of scores-2, scores-3, scores-4, and scores-5, 
and got 0.3098 minDCF and 5.076 \% ERR for Fixed data track, 
which outperformed the baseline by 1.3 \% minDCF and 2.2 \% ERR respectively.
\begin{table*}[b]
  \caption{Validation results.}
  \label{tab:example}
  \centering
  \begin{tabular}{lllllllll}
    \toprule
    \textbf{Score file} & \textbf{Topology} & \textbf{Train frame} & \textbf{Eval. fram} & \textbf{Augment.}
    & \multicolumn{2}{c}{\textbf{Vox1 Set}} & \multicolumn{2}{c}{\textbf{Vox1-H Set}}\\
    & & & & & minDCF & ERR /\% & minDCF & ERR /\% \\
    \midrule
    Scores-1   & Fast ResNet-34 & 2 s & 4 s & No & 0.1609 & 2.047 & 0.2459 & 3.925 \\
    Scores-2   & Half ResNet-34	& 2 s	& 4 s	& No & 0.1170 & 1.585 & 0.2168 & 3.436 \\
    Scores-3   & Half ResNet-34 & 2 s & 0   & No & 0.1157 & 1.668 & 0.2156 & 3.495 \\
    Scores-4   & Half ResNet-34 & 2 s & 4 s & Yes& 0.0864	& 1.180 & 0.1563 & \textbf{2.512} \\
    Scores-5   & Half ResNet-34 & 2 s & 0   & Yes& 0.0985 & 1.311 & 0.1651 & 2.744 \\
    fusion-avg & Scores-2,3,4,5 & N/A & N/A & N/A& 0.0820 & 1.251 & 0.1628 & 2.670 \\
    fusion-opt & Scores-2,3,4,5 & N/A & N/A & N/A& 0.0730 & 1.220 & 0.1474 & \textbf{2.463} \\
    \bottomrule
  \end{tabular}
  
\end{table*}

% conclusion
\section{Conclusions}
This is the first year we participate VoxCeleb Speaker Challenge, 
and we learn a lot from this challenge, especially from the open source trainer. However, 
since we are totally layman to this field, we have limited capability to improve the existing model. 
Even though our final rank is quite unsatisfactory, we will keep learning and keep improving 
because it is just a start. Hoping next year, we can get higher rank and make more contributions.

% acknowledgements
\section{Acknowledgements}
Thanks to VoxCeleb, thanks to the organizers including J. S. Chung, A. Nagrani, etc. 
You know, your pioneering and selfless works has significant impact on speaker recognition. 
Meanwhile, it is also very beneficial to those who are interested in this field, 
especially for freshman like me. 
I hope VoxCeleb can be more and more popular and widely used in the future.   
This challenge is just a start for me, 
I am on my way to make some contributions to speaker recognition from now on. 
Even though there are so many uncertainties in front, 
I have the courage and confidence to think, to explore to contribute. 
I just want to use myself as an example to tell this world, 
no matter where you are from, no matter how old you are, if you have interest on it, pick it up. 
What's more, don't worry, just run.

% reference
% \section{References}   
\bibliographystyle{IEEEtran}

\bibliography{reference} 

\end{document}